\documentclass[aps,prd,showpacs,showkeys,preprint,secnumarabic]{revtex4-1}
\usepackage{amsmath}
\usepackage{tikz}

\begin{document}
\title {Inhomogeneous and Homogeneous Renormalization Group Equations for the Effective Potential}

\author{H. I. Alrebdi$^{1}$}\email{haifa.alrebdi@gmail.com}
\author{H. A. Alhendi$^{1,2}$}\email{alhendi@ksu.edu.sa}
\author{T. Barakat$^{1}$}\email{tbarakat@ksu.edu.sa}
\affiliation{$^{1}$Department of Physics and Astronomy, College of Science, King Saud University, P. O. Box 2455, Riyadh 11451, Saudi Arabia}
\affiliation{$^{2}$NSRI, King Abdullaziz City of Science and Technology, P. O. Box 6086, Riyadh 11442, Saudi Arabia}
\date{\today}

\begin{abstract}
The inhomogeneous renormalization group equation for the effective potential is rederived. It is shown that when the effective potential is normalized by the normalization condition on the generating functional, its renormalization group equation is homogeneous. This is demonstrated in the case of massive $\phi^4$-theory. We also show that for the case of spontaneous symmetry breaking, the normalized effective potential is completely different from the symmetric case, though the two cases satisfy the same RGE with the same RG-functions. It is concluded that the vacuum energy density arises only in the case of SSB.
\end{abstract}

\pacs{11.10.Gh, 11.15.Bt, 11.30.Qc}
\keywords{Effective potential, Renormalization group equation, Normalization generating functional}

\maketitle

\section{Introduction:}
It is now generally recognized that the effective potential (EP) of a renormalizable field theory is an essential tool in the investigation of the vacuum structure and stability of the theory at zero and finite temperature. For weak coupling one uses\cite{b1} the perturbative loop expansion. As was emphasized in ref.\cite{b1} , the region of validity of the loop expansion may be appreciably extended using the renormalization group equation (RGE) satisfied by the EP. The RGE of the EP is also useful in determining its asymptotic behavior \cite{b2,b3}and in facilitating calculation of higher-loop contributions\cite{b5} and obtaining conditions for their validity \cite{b4}.

There has been much recent interest in the RGE of the EP V$(\phi)$ of a massive scalar field theory and its relation to the vacuum energy. Several authors \cite{b6,b7,b8,b9,b10,b11,b12} assert that, in order to satisfy the RGE, a $\phi$-independent vacuum energy term should be added to the Lagrangian density in the massive case with a corresponding RG- $\beta$ function included. Such an assertion was to our knowledge, first made in ref.\cite{b6}.

Those authors who find it necessary to add an additional term to the Lagrangian density, for compatibility with the RGE of the EP, invariably assume that this \emph{RGE is homogenous}. Our finding is that, in fact, for the massive case the \emph{RGE is generally inhomogeneous}. The inhomogeneity, is the consequence of the non-multiplicative renormalization required for the vacuum graphs. This fact has been emphasized in ref.\cite{b13}. In the context of the EP the inhomogeneity of the RGE has clearly been stated in ref.\cite{b14} and several ways for circumventing it and reducing the RGE to the homogeneous form have been discussed.

Some authors assume that the addition of a $\phi$-independent term to the tree level potential $V_{0}(\phi)$ to be equivalent to inclusion of the vacuum energy that requires an extra vacuum $\beta$- function\cite{b6,b7,b8}. One notes, however, that the generation of vacuum graphs, and a corresponding vacuum energy contribution to the EP, does not require the introduction of such a term in the tree level potential $V_{0}(\phi)$, since it is to be generated from radiative corrections under the normalization condition of the generating functional of the effective action. It is thus expected that the EP of a renormalizable field theory to satisfy the RGE \emph{without need for any modification} since this is a necessary condition of renormalizability. This observation is our main motivation in seeking to investigate this issue by making explicit calculations. 

In section II we derive the RGE for a massive scalar theory and show it to be, in general, inhomogeneous as stated in ref.\cite{b14}. In section III we discuss the functional formulation and show that the normalization condition $W[0]=1$ on the generating functional W[J] induces a particular normalization on the EP. This normalization removes the contribution of the vacuum graphs and leads to a homogeneous RGE for the EP. It also distinguishes the case of spontaneous symmetry breaking (SSB) from the normal case. In section IV we construct the normalized EP and demonstrate that, for both the normal case and that SSB, the homogeneous RGE is satisfied to the given order in the loop expansion with the same renormalization group functions. These constructions clarify the ambiguities that have afflicted the issue of an additional term in $V_{0}(\phi)$ for compatibility with the RGE. In section V we apply our construction to massive $\phi^{4}$-theory and show that although the expression of the effective potential of the symmetric case is completely different from the asymmetric case they satisfy the same RGE with the same RG-functions. In section VI we present our conclusion.

\section{Renormalization Group Equation for the Effective Potential:}
The EP of a renormalizable scalar field theory is given in terms of the renormalized 1PI Green's functions $\Gamma^{(n)}(m,\lambda,\mu)$, at zero external momenta, by
\begin{equation}
V(\phi)=-\sum\limits_{n=0}^\infty\frac{1}{n!}\Gamma^{(n)}\phi^{n}
\label{eq:h_1}
\end{equation}
The RGE for $V(\phi)$ is a consequence of the RGE for $\Gamma^{(n)}$. For $n\neq0$, the RGE of $\Gamma^{(n)}$ in the mass-independent scheme is the standard homogeneous equation.
\begin{equation}
\left\{{\mu\frac{\partial}{\partial \mu}+\beta}\frac{\partial}{\partial \lambda}+m\gamma_{m}\frac{\partial}{\partial m}+n\gamma\right\}\Gamma^{(n)}(m,\lambda,\mu)=0,\;\;\;\;\; n\neq0
\label{eq:h_2}
\end{equation}
where $\beta, \gamma_{m}$ and $\gamma$ are the renormalization group functions.
\\\\This renormalizability constraint follows from the fact that these $\Gamma^{(n)}$ are multiplicatively renormalizable:
\begin{equation}
\Gamma^{(n)}(m,\lambda,\mu)=\lim_{\epsilon\to 0} Z^{\frac{n}{2}}	\left(\lambda_o\mu^{-\epsilon}, \epsilon\right)\Gamma^{(n)}_{o}(\lambda_o, m_o, \epsilon),\;\;\;\;\; n\neq0
\label{eq:h_3}
\end{equation}
where $\lambda_o=\mu^{\epsilon}Z_\lambda\lambda$,  \;$m^{2}_{o}=m^{2}Z_m$ and $Z, Z_\lambda, Z_m$ are the renormalization constants. For the massive case, $\Gamma^{(o)}$ must, however, be treated separately, since $\Gamma^{(o)}_{o}$ is not multiplicatively renormalizable\cite{b13}. This is due to the fact that $\lim_{\epsilon\to 0}\Gamma^{(o)}_{o}(\lambda_o, m_o, \epsilon)$ does not exist. Consider, for example, the $\lambda$-independent contribution to $\Gamma^{(n)}_{o}$ in massive $\lambda\phi^{4}$ theory. This is given by the circle diagram which, for $\lambda=0$, gives

\begin{equation}\begin{aligned}
\tikz\draw[thick]circle (0.3);
&=-\frac{1}{2(2\pi)^{n}}\int\ln(\frac{p^{2}+m^{2}_{o}}{\mu^{2}})d^{n}p=\frac{1}{2}\mu^{-\epsilon}\frac{m^{4}}{(4\pi)^{2}}(\frac{m^{2}}{4\pi\mu^{2}})^{-\frac{\epsilon}{2}}\;\Gamma(\frac{\epsilon}{2}-2)\\
&=\frac{m^{4}}{64\pi^{2}}(-\ln\frac{m^2}{\mu^2}+\ln4\pi+\frac{3}{2}-\gamma_E+\frac{2}{\epsilon}+....)
\end{aligned}
\label{eq:h_4}
\end{equation}

This can only be renormalized by adding an arbitrary additive renormalization factor that removes the pole in $\epsilon$ and leaves an undetermined finite part. This indeterminacy persists to all orders in perturbation so that the vacuum contribution is in effect an arbitrary $\phi$-independent term. We shall now see that this term does not affect the RGE of the EP since it cancels on both sides of the resulting inhomogeneous RGE. This equation is obtained from equation (\ref{eq:h_1}) and (\ref{eq:h_2}) which immediately yield:
\begin{equation}
\left\{{\mu\frac{\partial}{\partial \mu}+\beta}\frac{\partial}{\partial \lambda}+m\gamma_{m}\frac{\partial}{\partial m}+\phi\gamma\frac{\partial}{\partial\phi}\right\}V(\phi)=C
\label{eq:h_5}
\end{equation}
where $C(\lambda,m,\mu)$ is given by
\begin{equation}
C=-(\mu\frac{\partial}{\partial \mu}+\beta\frac{\partial}{\partial \lambda}+m\gamma_{m}\frac{\partial}{\partial m})\Gamma^{(o)}
\label{eq:h_6}
\end{equation}
The presence of the inhomogeneous term in (\ref{eq:h_5}) is clearly to ensure that the arbitrary vacuum contribution to the EP is canceled in the RGE and does not affect any physical consequences. It is therefore desirable to explicitly subtract this contribution and rewrite equations (\ref{eq:h_5}) and (\ref{eq:h_6}) in the form:
\begin{equation}
\left\{{\mu\frac{\partial}{\partial \mu}+\beta}\frac{\partial}{\partial \lambda}+m\gamma_{m}\frac{\partial}{\partial m}+\phi\gamma\frac{\partial}{\partial\phi}\right\}\left\{V(\phi)-V(o)\right\}=0
\label{eq:h_7}
\end{equation}
assuming the $V(o)$ is finite. This equation has previously been obtained in\cite{b14}, where $V(o)$ is interpreted as a contribution to the vacuum energy upon which no observable can depend outside of gravity.

Although the presence of an arbitrary additive $\phi$-independent term in $V(\phi)$ dose not affect the RGE of the EP, it is desirable that such term be fixed by normalizing the EP in a definite manner for specific calculations. We show in the following section that the usual normalization\cite{b15} of the generating functional in the Feynman path-integral formulation induces definite normalization for the EP that fixes its value at any particular value of the field.

\section{Normalization of the Effective Potential:}
The generating functional $W[J]$ for a scalar field theory, normalized such that $W[0]=1$, is given by
\begin{equation}
W[J]=\frac{\int D\phi\exp \left[\frac{i}{\hbar}\left\{S[\phi]+\hbar(J,\phi)\right\}\right]}{\int D\phi\exp\left\{\frac{i}{\hbar}S[\phi]\right\}}
\label{eq:3_1}
\end{equation}
where $S[\phi]$ is the action, and
\begin{equation}
(J,\phi)=\int d^{4}x J(x) \phi(x)
\label{eq:3_2}
\end{equation}
The classical field $\phi_{c}(x)$ and the effective action $\Gamma[\phi_{c}]$ are given by:
\begin{equation}
\phi_{c}(x)=\frac{1}{i}\frac{\delta}{\delta J(x)}\ln\stackrel{}{}W[J]
\label{eq:3_3}
\end{equation}
\begin{equation}
\Gamma[\phi_{c}]=\frac{\hbar}{i} \ln W[J]-\hbar(J,\phi_{c})
\label{eq:3_4}
\end{equation}
where in the last equation, $J(x)$ is obtained from (\ref{eq:3_3}) in terms of $\phi_{c}(x)$.\\

Equations (\ref{eq:3_3}) and (\ref{eq:3_4}) imply
\begin{equation}
\frac{\delta\Gamma[\phi_{c}] }{\delta\phi_{c}(x)}+\hbar J(x)=0
\label{eq:3_5}
\end{equation}
In terms of the effective action, the EP $V(\eta)$ is obtained by taking for $\phi_{c}(x)$ a constant field $\eta$:
\begin{equation}
(2\pi)^{4} \delta^{(4)}(o) V(\eta)=-\Gamma(\phi_{c}=\eta)
\label{eq:3_6}
\end{equation}
The steepest-descent evaluation of $W[J]$, leading to the loop expansion for the EP, is effected by expanding $S[\phi]+\hbar(J,\phi)$ about the field $\phi_{o}(x)$ which minimizes it:

\begin{equation}
\frac{\delta S [\phi] }{\delta\phi(x)}|_{\phi=\phi_{o}}+\hbar J(x)=0
\label{eq:3_7}
\end{equation}

\begin{equation}
S[\phi]+\hbar(J,\phi)= S[\phi_{o}]+\hbar(J,\phi_{o})+\frac{1}{2}\int\int d^{4}x d^{4}y\left(\phi-\phi_{o}\right)(x)\left\{\frac{\delta^{2}S[\phi]}{\delta\phi(x)\delta\phi(y)}\right\}|_{\phi=\phi_{o}}(\phi-\phi_{o})(y)+...
\label{eq:3_8}
\end{equation}
Functional integration over $\phi$ enables one to replace $\phi-\phi_{o}$ by $\phi$ in the quadratic term. When this is applied to both the numerator and dominator in equ.(\ref{eq:3_1}) one obtains:

\begin{equation}
W[J]=\frac{e^{{\frac{i}{\hbar}}\left\{S[\phi_{o}]+\hbar(J,\phi_{o})\right\}}\int D\phi\exp\frac{-i}{2\hbar}\left\{\int\int d^{4}x d^{4}y\phi(x)\widehat{A}(\phi_{o})\phi(y)+...\right\} }{e^{{\frac{i}{\hbar}}S[v]}\int D\phi\exp\frac{-i}{2\hbar}\left\{\int\int d^{4}x d^{4}y\phi(x)\widehat{A}(v)\phi(y)+...\right\}}               
\label{eq:3_9}
\end{equation}
where $v$ and $\widehat{A}$ are defined by

\begin{equation}
\frac{\delta S [\phi] }{\delta\phi(x)}|_{\phi=v}=0,\stackrel{}{}\stackrel{}{}\stackrel{}{}\stackrel{}{}\widehat{A}(\phi)=\frac{\delta^{2} S [\phi] }{\delta\phi(x)\delta\phi(y)}
\label{eq:3_10}
\end{equation}
For the Lagrangian density
\begin{equation}
\mathcal{L}=\frac{1}{2}\partial_{\mu}\phi\partial^{\mu}\phi-\frac{1}{2}m^{2}\phi^{2}-V_{o}(\phi)
\label{eq:3_11}
\end{equation}
one has
\begin{equation}
\widehat{A}(\phi)=(\partial_{\mu}\partial^{\mu}+m^{2}+V^{''}_{o}(\phi))\delta^{4}(x-y)
\label{eq:3_12}
\end{equation}
As well-known\cite{b16} the solution $\phi_{o}=\phi_{o}(J)$ to equ.(\ref{eq:3_7}) is $\phi_{o}=\phi_{c}$ to first order in $\hbar$. This order is sufficient for the calculation of $W[J]$ to order $\hbar^{2}$.

Evaluation of the integrals in (\ref{eq:3_9}) is performed in Euclidean space where one obtains:

\begin{equation}
W[J]=\exp\left\{-\frac{1}{\hbar}\left(S_{E}[\phi_{c}]+\hbar(J,\phi_{c})-S[v]\right)\right\}\left(\frac{\det\stackrel{}{}\widehat{A}_{E}(\phi_{c})}{\det\stackrel{}{}\widehat{A}_{E}(v)}\right)^{-\frac{1}{2}}+...
\label{eq:3_13}
\end{equation}
For $\phi_{o}=\eta$, a constant field,
\begin{equation}
S_{E}(\eta)=(\frac{1}{2}m^{2}\eta^{2}+V_{o}(\eta))(2\pi)^{4}\delta^{(4)}(o)
\label{eq:3_14}
\end{equation}
In calculating $\Gamma(\eta)$ using (\ref{eq:3_4}) and (\ref{eq:3_13}) the dependence on $J$ cancels so that

\begin{equation}
\Gamma_{E}(\eta)=(\frac{1}{2}m^{2}\eta^{2}+V_{o}(\eta)-\frac{1}{2}m^{2}v^{2}+V_{o}(v))(2\pi)^{4}\delta^{(4)}(o)+\frac{\hbar}{2}\left(\ln\det\widehat{A}_{E}(\eta)-\ln\det\widehat{A}_{E}(v)\right)+...
\label{eq:3_15}
\end{equation}
where
\begin{equation}
\widehat{A}_{E}(\eta)=(-\partial_{\mu}\partial_{\mu}+m^{2}+V^{''}_{o}(\eta))
\label{eq:3_16}
\end{equation}
For the $\hbar$-term in (\ref{eq:3_15}) one finds\cite{b16}

\begin{equation}
\ln\det\widehat{A}_{E}(\eta)=\frac{1}{32\pi^{2}}(m^{2}+V^{''}_{o}(\eta))^{2}\left[\ln\frac{m^{2}+V^{''}_{o}(\eta)}{\mu^{2}}-\frac{3}{2}\right](2\pi)^{4}\delta^{(4)}(o)
\label{eq:3_17}
\end{equation}

where $\mu$ is an arbitrary parameter of mass-dimension. Then the Euclidean version of (\ref{eq:3_6}) finally gives for the EP $V(\eta)$:

\begin{equation}
V(\eta)=\frac{1}{2}m^{2}\eta^{2}+ V_{o}(\eta)+\frac{\hbar}{64\pi^{2}}(m^{2}+V^{''}_{o}(\eta))^{2}\left\{\ln\frac{m^{2}+V^{''}_{o}(\eta)}{\mu^{2}}-\frac{3}{2}\right\}-(\eta\rightarrow v)+O(\hbar^{2})
\label{eq:3_18}
\end{equation}
\\
Thus one conclude that the normalization of the generating functional $W[J]$ as in equ.(\ref{eq:3_1}) directly leads to the particular expression (\ref{eq:3_18}) for the effective potential. This expression completely specifies the value of $V(\eta)$ at any value of the field and does not allow an arbitrary additive constant.

Now the value $v$ of the field at which $V(\eta)$ is subtracted in (\ref{eq:3_18}), due to the normalization of $W[J]$, has the physical interpretation that follows from the first of equ.(\ref{eq:3_10}). This equation indicates that $\eta=v$ is an extremum of the action. To generate a suitable perturbative series this extremum should be taken as the absolute minimum of the action $S[\phi]$, i.e. the physical vacuum of the theory. In terms of Feyman graphs this amounts to subtraction of the contribution of the vacuum graphs from the theory. Thus the expression in (\ref{eq:3_18}) receives no contribution from these graphs.

We obtain, in the following section, the RGE for the EP normalized by the condition $W[0]=1$.

\section{Renormalization Group Equation for the Normalized Effective Potential:}
We first prove that the normalization condition $W[0]=1$ implies that the EP is in general, and not just perturbatively, normalized so that  
\begin{equation}
V(v)=0
\label{eq:4_1}
\end{equation}
where $v$ is obtained from the tadpole condition $\left(\frac{\partial V}{\partial \eta}\right)_{\eta=v}=-\Gamma^{(1)}(0)=0$\\\\
Towards this purpose one writes the generating functional $W[J]$ in the form
\begin{equation}
W[J]=\frac{\widetilde{W}[J]}{\widetilde{W}[0]}
\label{eq:4_2}
\end{equation}
so that $\Gamma[\phi_{c}]$ is of the form
\begin{equation}
\Gamma[\phi_{c}]= G[J]-G[0]
\label{eq:4_3}
\end{equation}
For $\phi_{c}(x)=\eta$ , a constant field, one has $J(x)=j$ where\\ 
\begin{equation}
\left(\frac{\delta \Gamma [\phi_{c}]}{\delta \phi_{c}(x)}\right)_{\phi_{c}(x)=\eta}+\hbar j=0
\label{eq:4_4}
\end{equation}
For the spicial case $j=0$ the corresponding constant field is $\phi=v$ where 
\begin{equation}
\left(\frac{\delta \Gamma [\phi_{c}]}{\delta \phi_{c}(x)}\right)_{\phi_{c}(x)=v}=0
\label{eq:4_5}
\end{equation}
Equations (\ref{eq:4_3}), (\ref{eq:4_4}) and (\ref{eq:4_5}) show that $\Gamma(\eta)$ is of the form
\begin{equation}
\Gamma(\eta)=g(\eta)-g(v)
\label{eq:4_6}
\end{equation}
where $g(\eta)=G(j(\eta))$ and $j(\eta=v)=0$ . Equ.(\ref{eq:4_6}), using equ.(\ref{eq:3_6}), directly leads to the condition (\ref{eq:4_1}) for the EP while (\ref{eq:4_5}) yields the condition $\left(\frac{\partial V(\eta)}{\partial \eta}\right)_{\eta=v}=0$ as previously stated.

To obtain the RGE for the normalized EP we use equ.(\ref{eq:h_7}) under the conditions $V(v)=0$ , $\left(\frac{\partial V}{\partial \eta}\right)_{\eta=v}=0$.\\\\
Setting $\phi=v$ in equ.(\ref{eq:h_7}), one then obtains 
\begin{equation}
DV(0)=0
\label{eq:4_7}
\end{equation}
where
\begin{equation}
D=\mu\frac{\partial}{\partial\mu}+\beta\frac{\partial}{\partial\lambda}+\gamma_{m} m\frac{\partial}{\partial m}
\label{eq:4_8}
\end{equation}
Thus, trivially, one could write
\begin{equation}
\left(D+\phi\gamma\frac{\partial}{\partial\phi}\right)V(0)=0
\label{eq:4_9}
\end{equation}
and one finally finds for the normalized EP the RGE
\begin{equation}
\left(\mu\frac{\partial}{\partial\mu}+\beta\frac{\partial}{\partial\lambda}+m\gamma_{m} \frac{\partial}{\partial m}+\phi\gamma\frac{\partial}{\partial\phi}\right)V(\phi)=0
\label{eq:4_10}
\end{equation}\\
One thus concludes that with the normalization of the EP induced by $W[0]=1$ for the generating functional, the RGE for the EP is homogeneous. This important conclusion has not, to our knowledge, been previously explicitly noted in the literature.

For the case of spontaneous symmetry breaking (SSB), the extremum condition $\left(\frac{\partial V}{\partial\phi}\right)_{\phi=v}=0$, is satisfied by the vacuum value $v\neq 0$, whereas in the normal case $v=0$. In both cases the RGE for the normalized EP is the homogeneous equ.(\ref{eq:4_10}), and the normalization condition $W[0]=1$ requires subtraction of the EP at its minimum $\phi=v$ which is calculated from the tadpole condition\cite{b17,b18}. Thus our analysis completely removes the ambiguities concerning the form of the RGE satisfied by the EP.

\section{Application to Massive $\phi^{4}$ Theory:}

For the case where the physical vacuum is at $\phi=0$ one has $V_{o}=\frac{1}{2}m^{2}\phi^{2}+\frac{\lambda}{24}\phi^{4}$ with $m^{2}>0$, so that $v=0$. The normalized EP is then 

\begin{equation}
V(\phi)=V_{o}(\phi)+\frac{\hbar}{64\pi^{2}}M^{4}(\phi)\left(\ln \frac{M^{2}(\phi)}{\mu^{2}}-\frac{3}{2}\right)-\frac{\hbar}{64\pi^{2}}m^{4}\left(\ln \frac{m^{2}}{\mu^{2}}-\frac{3}{2}\right)+...
\label{eq:5_1}
\end{equation}
where
\begin{equation}
M^{2}(\phi)=m^{2}+\frac{1}{2}\lambda\phi^{2}
\label{eq:5_2}
\end{equation}\\
This expression for $V(\phi)$ satisfies the RGE (\ref{eq:4_10}), to first order in $\hbar$. This may be checked using the one-loop renormalization group functions:

\begin{equation}
\beta=\frac{3\hbar\lambda^{2}}{16\pi^{2}},  \stackrel{}{} \gamma_{m}=\frac{\hbar\lambda}{32\pi^{2}}, \stackrel{}{} \gamma=0
\label{eq:5_3}
\end{equation}\\

For the case of SSB with $V^{(S)}_{o}=-\frac{1}{2}m^{2}\phi^{2}+\frac{\lambda}{24}\phi^{4}$ and $m^{2}>0$, the position of the minimum is at $\phi=v$, where $v$ to the first order in $\hbar$ is

\begin{equation}
v^{2}=\frac{6m^{2}}{\lambda}-\frac{3\hbar m^{2}}{8\pi^{2}}\left(\ln \frac{2m^{2}}{\mu^{2}}-1\right)
\label{eq:5_4}
\end{equation}
The normalized EP $V(\phi)$ is then:

\begin{equation}
V(\phi)=V^{(S)}_{o}(\phi)+\frac{\hbar}{64\pi^{2}}M^{4}_{S}(\phi)\left(\ln \frac{M^{2}_{S}(\phi)}{\mu^{2}}-\frac{3}{2}\right)+\frac{3m^{4}}{2\lambda}-\frac{\hbar m^{4}}{16\pi^{2}}\left(\ln \frac{2m^{2}}{\mu^{2}}-\frac{3}{2}\right)
\label{eq:5_5}
\end{equation}\\
where
\begin{equation}
M^{2}_{S}=-m^{2}+\frac{1}{2}\lambda\phi^{2}
\label{eq:5_6}
\end{equation}
One may now check, using the same RG-functions (\ref{eq:5_3}), that this expression for $V(\phi)$ also satisfies the homogeneous RGE (\ref{eq:4_10}) to order $\hbar$. Thus equ.(\ref{eq:4_10}) is valid for both cases, the symmetric case and the one with SSB, although the expressions for the normalized EP are different for these two cases.

This distinction between the two cases, and the relevance to the RGE for the normalized potential, has not, to our knowledge, been noted before.
The term containing $\frac{m^{4}}{\lambda}$, in equ.(\ref{eq:5_5}) arises only in the case of SSB where it does not introduce a singularity since the expansion is, in this case, about $\phi^{2}=v^{2}$ for $V(\phi)$. This is in contrast to the work of other authors\cite{b6,b7,b8,b10,b14} where a term in $\frac{m^{4}}{\lambda}$ is added to $V_{o}(\phi)$ so as to satisfy a homogeneous RGE. In our case additional terms such as those in equ.(\ref{eq:5_5}) will automatically arise in the case of SSB for any renormalizable $V_{o}(\phi)$ while the RGE for the normalized EP is always the homogeneous equation in (\ref{eq:4_10}).

One finally notes that to obtain a homogeneous RGE for the EP one may subtract at any finite $\phi=\phi_{o}$. This will not, however, lead to an EP that satisfies the normalization condition of the generating functional. In the case of the standard model, subtraction at $\phi=0$, as has been used for example in ref.\cite{b19}, produces a complex expression for the EP.

\section{Conclusions:}
In this paper the inhomogeneous RGE has been derived. We then obtained the general form of the EP under the usual normalization of the generating functional in the path-integral formulation. This normalized EP distinguishes between the cases where the vacuum occurs at a zero or non-zero value of $\phi$. The RGE for the normalized EP is homogeneous in both cases, with the same RG functions. These results are illustrated in the case of massive $\phi^{4}$ theory to one-loop order. It shown that additive terms arise only in the case of SSB.
Finally we note that our work may be extended to the case of gauge fields and to finite-temperature field theory.


\begin{thebibliography}{99}

\bibitem{b1} S. Coleman and E. Weinberg, Phys. Rev. D7, 1888 (1973).
\bibitem{b2} M. O. Taha, Phys. Rev. D26, 2766 (1982).
\bibitem{b3} H. Alhendi and M. O. Taha, Phys. Lett. B300, 373 (1993).
\bibitem{b4} H. Alhendi, Phys. Rev. D37, 3749 (1988); ibid D40, 683(E) (1989).
\bibitem{b5} H. A. Alhendi, T. Barakat and I. Gh. Loqman, Phys. Rev. D82, 053008 (2010).
\bibitem{b6} Y. Fujimoto, L. O Raifeartaigh and G. Parravicini, Nucl. Phys. B212, 268 (1983).
\bibitem{b7} B. Kastening, Phys. Lett. B283, 287 (1992); Phys. Rev. D54, 3965 (1996); Phys. Rev. D57, 3567 (1998).
\bibitem{b8} M. Bando, T. Kugo, N. Maekawa and H. Nakano, Phys. Lett. B301, 83 (1993).
\bibitem{b9} H. Nakano and Y. Yoshida, Phys. Rev. D49, 5393 (1994).
\bibitem{b10} C. Ford, Phys. Rev. D50, 7531 (1994).
\bibitem{b11} M. B. Einhorn and D. R. T. Jones, JHEP. 0704, 051 (2007).
\bibitem{b12} J. Sola, J. Phys. Conf. Ser. 453, 012015 (2013).
\bibitem{b13} J. Zinn-Justin, “Quantum Field Theory and Critical Phenomena” Clarendon Press, Oxford (1989).
\bibitem{b14} C. Ford, D. R. T. Jones, P. W. Stephenson and M. B. Einhorn, Nucl. Phys. B395, 17 (1993).
\bibitem{b15} See, for example, C. Itzykson and J. B. Zuber, “Quantum Field Theory”, McGraw-Hill (1980).
\bibitem{b16} See, for example, P. Ramond, “Field Theory: A Modern Primer”, Addison-Wesley (1990).
\bibitem{b17} S. Weinberg, Phys. Rev. D7, 2887 (1973).
\bibitem{b18} S. Y. Lee and A. M. Sciaccaluga, Nucl. Phys. B96, 435 (1975).
\bibitem{b19} M. J. Duncan, R. Philippe and M. Sher, Phys. Lett. B153, 165 (1985).
\end{thebibliography}
\end{document}